\documentclass[12pt,a4paper,dvips]{article}
\usepackage{epsfig,wrapfig} 
\renewcommand{\section}[1]{\vspace{6pt} \noindent\mbox{#1} \newline \noindent}
\renewcommand{\subsection}[1]{\vspace{6pt} \noindent\mbox{\underline{#1}} 
\newline \noindent}
\renewcommand{\subsubsection}[1]{\vspace{6pt} \noindent\mbox{\underline{#1}}
\noindent}

\newfont{\sansb}{cmssbx10}
\newfont{\sans}{cmss10}

\begin{document}
{\small OG 4.3. \vspace{-24pt}\\}     
{\center \LARGE{ TeV Gamma-ray Observations of Southern AGN with the CANGAROO 3.8m Telescope}
\vspace{6pt}\\}

 M. D. ROBERTS,$^{1}$ S. A. DAZELY,$^{2}$ P. G. EDWARDS,$^{3}$
 T. HARA,$^{4}$ A. HAYAMI,$^{5}$ S. KAMEI,$^{5}$ T. KIFUNE,$^{1}$
 R. KITA,$^{6}$ T. KONISHI,$^{7}$ A. MASAIKE,$^{8}$ Y. MATSUBARA,$^{9}$
 Y. MIZUMOTO$^{10}$, M. MORI$^{11}$, H. MURAISHI,$^{6}$ Y. MURAKI,$^{9}$ 
K. NISHIJIMA,$^{12}$ S. OGIO,$^{5}$ J. R. PATTERSON,$^{2}$ G. P. ROWELL,$^{2}$
 T. SAKO,$^{9}$ K. SAKURAZAWA,$^{5}$ R. SUSUKITA,$^{13}$ A. SUZUKI,$^{7}$
 R. SUZUKI,$^{5}$ T. TAMURA,$^{14}$ T. TANIMORI,$^{5}$ G. J. THORNTON,$^{1}$
 S. YANAGITA,$^{6}$ T. YOSHIDA,$^{6}$ T. YOSHIKOSHI$^{1}$ 

{\it{$^{1}$Institute for Cosmic Ray Research, University of Tokyo, 
Tokyo 188, Japan.\\
$^{2}$Department of Physics and Mathematical Physics, University of Adelaide, 
South Australia 5005, Australia.\\
$^{3}$Institute of Space and Astronomical Science, Sagamihara 152, Japan.\\
$^{4}$Faculty of Commercial Science, Yamanashi Gakuin University, Kofu 400, Japan.\\
$^{5}$Department of Physics, Tokyo Institute of Technology, Tokyo 152, Japan.\\
$^{6}$Faculty of Science, Ibaraki University, Mito 310, Japan.\\
$^{7}$Department of Physics, Kobe University, Hyogo 637, Japan.\\
$^{8}$Department of Physics, Kyoto University, Kyoto 606, Japan.\\
$^{9}$Solar-Terrestrial Environment Laboratory, Nagoya University, Nagoya 464, Japan \\ 
$^{10}$National Astronomical Observatory, Tokyo 181, Japan.\\
$^{11}$Faculty of Education, Miyagi University of Education, Sendai 980, Japan.\\
$^{12}$Department of Physics, Tokai University, Hiratsuku 259, Japan. \\
$^{13}$Institute of Physical and Chemical Research, Wako, Saitama 351, Japan.\\
$^{14}$Faculty of Engineering, Kanagawa University, Yokohama 221, Japan.\\
}}

{\center ABSTRACT\\}

Since 1992 the CANGAROO 3.8m imaging telescope has been used to search for 
sources of TeV gamma-rays. Results are presented here for observations of
four Southern Hemisphere BL-Lacs - PKS0521-365 ,PKS2316-423, PKS2005-489  
and EXO0423-084. In addition to testing for steady DC emission, a night 
by night burst excess search has been performed for each source.

\vspace{6pt}

\setlength{\parindent}{1cm}

\section{{THE {\bf{CANGAROO}} IMAGING ATMOSPHERIC CHERENKOV TELESCOPE}}
The CANGAROO 3.8m imaging telescope is located near Woomera, South Australia
(longitude $137^{\circ}47'$E, latitude $31^{\circ}06'$S, 160m a.s.l).
The reflector is a single 3.8m diameter parabolic dish with F1 optics.
The imaging camera consists of a square packed array of 10mm $\times$ 10mm
Hamamatsu R2248 photomultiplier tubes. The tube centers are separated
by $0.18^{\circ}$, giving
a total field of view (side-side) of $\sim 3^{\circ}$. 
The current gamma-ray
energy threshold of the 3.8m telescope is estimated to be $\sim 1.0$ TeV. Prior
to mirror recoating in November 1996 (which includes all data presented
in this paper) the energy threshold was somewhat higher ($\sim$ 2TeV).
For a more detailed description of the 3.8m telescope see Hara et al. 1993.

\section{DATA SAMPLE}
PKS2316-423 (RA = $23^{h}19^{m}05.8$, Dec= $-42^{\circ}$06'48'', z= 0.055) is
a BL-Lac object associated with a parent elliptical galaxy. It has previously
been studied over a wide range of wavelengths. Emission from 
radio through to x-ray
wavelengths is consistent with synchrotron radiation from electrons with
$E > 10^{13}$eV (Crawford and Fabian 1994). PKS2316-423 is not detected
by the EGRET telescope on the CGRO.
The CANGAROO 3.8m telescope observed PKS2316-423 during July 1996, for
a total of 26 hours on-source data and 25 hours off-source data.

PKS0521-365 (RA=$05^{h}22^{m}57.8$, Dec.= $-36^{\circ}$27'03'',Z=0.055)
is a radio selected BL-Lac. It was first detected as a strong radio
source more than 30 years ago. 
  PKS0521-365 was viewed by the EGRET experiment on the CGRO during
1992 from May 14 to June 4.
 A point source, consistent with the position
of PKS0521-365 was detected at a statistical significance of 4.3 $\sigma$.
The integral source flux above 100MeV was $(1.8 \pm 0.5) \times 10^{-7}$
photons ${\rm cm}^{-2}{\rm s}^{-1}$.
The hardness of the EGRET photon spectrum and the proximity of the
source make PKS0521-365 a candidate source for detectable levels 
of TeV gamma-ray emission.
PKS0521-365 has been observed by the CANGAROO 3.8m telescope 
for three consecutive years between 1993 and
1995. The CANGAROO raw data set consists of 52 on/off pairs with a total of
89 hours of on-source and 84 hours of off-source data.

PKS2005-489 (RA=$20^{h}09^{m}25.4$, Dec=$-48^{\circ}$49'54'', z=0.071) is an
x-ray
selected BL-Lac. X-ray measurements of PKS2005-489 by EXOSAT show extremely
large flux variations on timescales of hours(Giommi et al. 1990)
. PKS2005-489 is not detected by EGRET.
We have observed PKS2005-489 during August 1993 and during August/September 
1994 obtaining 41 hours of on-source and 38 hours of off-source data.

EXO0423-084 (RA = $04^{h}25^{m}50.7$, Dec=$ -08^{\circ}$33'43'', z= 0.039)
was detected at x-ray wavelengths by EXOSAT. The associated galaxy,
MCG 01-12-005, is a radio source free of emission lines. This, along with 
the high X-ray luminosity ($>10^{43}$ ergs ${\rm s}^{-1}$) makes EXO0423-084
a candidate BL-Lac object (Giommi et al. 1991). If this source is a BL-Lac,
it would be the third closest such object known (after Mkn 421 and Mkn 501),
and the closest in the southern hemisphere.
This source was observed during October of 1996 for a total of 19 hours 
on-source and 18 hours off-source data.

\section{ANALYSIS}
\begin{figure}[t]
\vspace{100mm}
\includegraphics{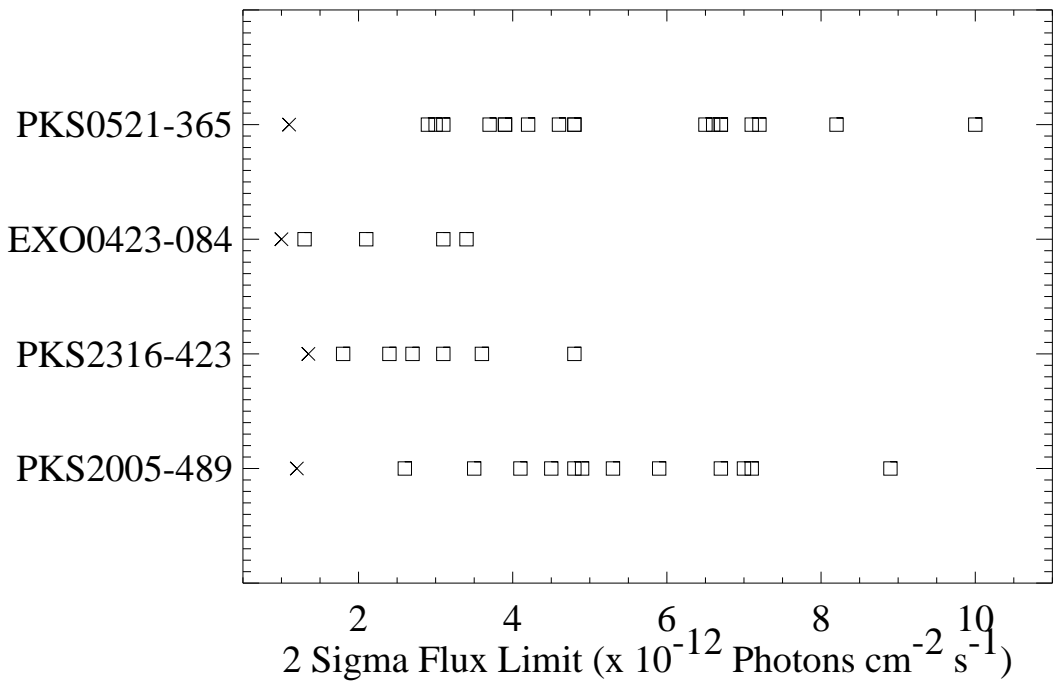}
\caption{Scatter plot of night by night $2\sigma$ flux limits for each of the
sources (indicated by squares). Also shown are the $2\sigma$ flux limits
for the total data set for each source (crosses). }
\label{flux}
\end{figure}

\begin{figure}[t]
\vspace{100mm}
\includegraphics{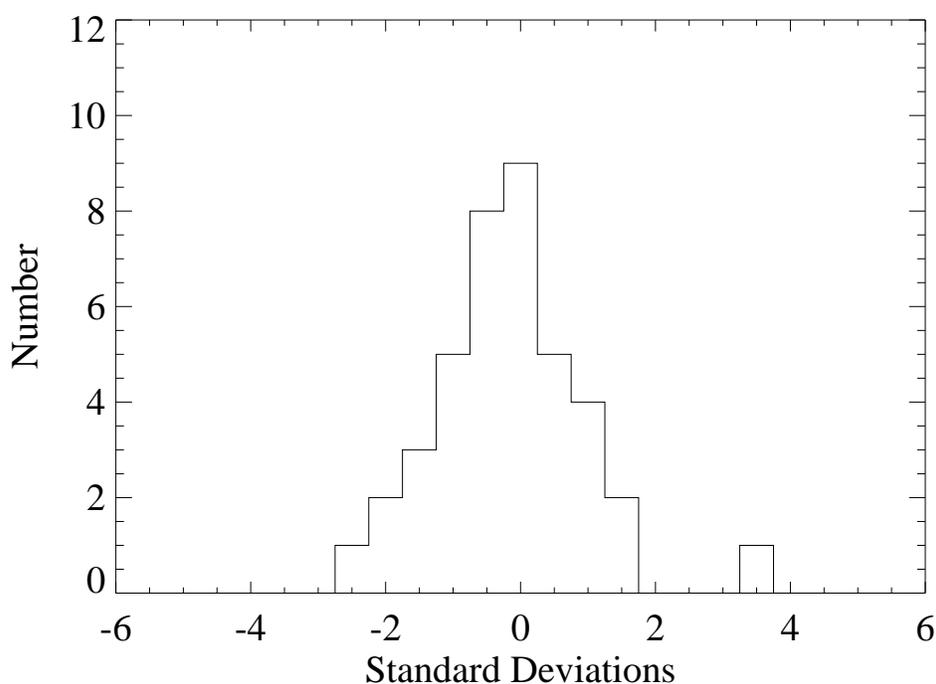}
\caption{Distribution of the significances of night by night
excesses for all sources.}
\label{sig}
\end{figure}

Prior to image analysis each data set is subjected to data 
integrity checks and photomultiplier gain/pedestal correction.
Images that are sufficiently large ($>$ 4 tubes and $>$ 20 photoelectrons)
are parameterized after Hillas (1985). 
The gamma-ray selection domains for the 3.8m telescope are defined
as\\
\\
$0.5^{\circ} <$ Distance $< 1.1^{\circ}$\\
$0.01^{\circ} <$ Width $< 0.08^{\circ}$\\
$0.1^{\circ}<$  Length $< 0.4^{\circ}$\\
$0.4 <$ Concentration $< 0.9$\\
Alpha $< 10^{\circ}$\\
\\
These cuts, based on Monte Carlo simulations, reject 98\% of
the background cosmic ray triggers while retaining 40\% of the gamma-rays.
For each source the total data set has been tested for gamma-ray emission. 
The significance of the on-source excesses have been estimated using
a method based on the maximum likelihood
method of Li and Ma (1983).
In the total data set for each source no significant excess is seen for
those events in the gamma-ray domain.
The calculated excesses are $-0.91 \sigma$ (PKS2005-489), $+0.22 \sigma$
(PKS2316-423), $-0.99 \sigma$ (EXO0423-084) and $-0.27 \sigma$
(PKS0521-365).
 Upper limits to steady DC emission have
been calculated after Protheroe (1984) and are shown in the scatter plot 
in figure~\ref{flux}.
We have also searched our data set for gamma-ray emission on a night by 
night basis. In general our observations of a source consist of a long
(several hours) on-source run, with a similar length off-source
 run, offset in RA to provide the same coverage of azimuth and zenith. 
The burst search has been performed by calculating the on-source excess
for each pair of on/off observations each night.
 In cases where there
is  no matching off-source run, an equivalent off-source run from another
night is used. Figure~\ref{sig}  shows the distribution of
 on-source excess significances
for all four sources. There is no evidence for gamma-ray bursts on
the timescale of $\sim$ 1 night for any of the sources. 
The most significant nightly excess (from PKS0521-365) has a nominal 
significance of $3.7 \sigma$ but after allowing for the number of 
searches performed this significance is reduced to less than 
$3\sigma$.
The $2\sigma$ upper limits to
gamma-ray emission for each observation are shown in figure~\ref{flux}.

\section{CONCLUSION}
Analysis of CANGAROO data shows no evidence for steady DC or burst emission
of TeV gamma-rays from the BL-Lacs PKS0521-365, PKS2005-489, PKS2316-423 and EXO0423-084.
 A sensible
comparison at TeV energies between these sources and the proven TeV sources
Mkn421 and Mkn501 depends upon the nature of the gamma-ray emission above
the threshold of CANGAROO. 
If we assume that the gamma-ray emission from Mkn421 and Mkn 501 extend
 up to 10TeV with an integral spectral power law of index -1.0, we can compare
the expected fluxes with the upper limits shown in figure~\ref{flux}.
 Under these assumptions the integral quiescent fluxes above 2TeV are:\\
Mkn421 F($>2$TeV) $\sim 2.1 \times 10^{-12}$ photons ${\rm cm}^{-2}{\rm s}^{-1}$
(for measurement taken from Dec. 1993-April 1994) Kerrick et al. 1995.\\
Mkn501 F($>2$TeV) $\sim 1.0 \times 10^{-12}$ photons ${\rm cm}^{-2}{\rm s}^{-1}$
(Catanese 1995)\\
If any of these sources reported in this paper were
capable of providing a steady DC flux at the level of Mkn 421 it would
be detectable by CANGAROO , albeit 
at low significance. We would not expect to see
a source of similar flux to Mkn501.
Observations by Whipple of Mkn421 have shown that it is capable of extremely
energetic bursts on timescales of hours to days. A burst with
F($>$250GeV) $\sim 4.0 \times 10^{-10}$ photons ${\rm cm}^{-2}{\rm s}^{-1}$
(F($>$2TeV) $\sim 4.0 \times 10^{-11}$ photons ${\rm cm}^{-2}{\rm s}^{-1}$
for a -1.0 spectral index and maximum photon energy of 10TeV) lasting
two hours would be easily detectable by CANGAROO at a significance of
around $13 \sigma$.

\section{ACKNOWLEDGEMENTS}
This work is supported by a Grant-in-Aid in Scientific Research from the
Japan Ministry of Education, Science and Culture, and also by the 
Australian Research Council and the International Science and 
Technology Program. MDR acknowledges the receipt of a JSPS fellowship from
the Japan Ministry of Education, Science and Culture.

\section{REFERENCES}
\setlength{\parindent}{-5mm}
\begin{list}{}{\topsep 0pt \partopsep 0pt \itemsep 0pt \leftmargin 5mm
\parsep 0pt \itemindent -5mm}
\vspace{-15pt}
\item Catanese, M. in Towards a Major Atmospheric Cherenkov
Detector 4, p348, 1995.
\item Crawford, C. S. and Fabian, A. C., Mon. Not. Royal Astron. Soc,
{\bf 266},669 ,1994.
\item Giommi, P. et al., Ap. J. {\bf 356}, 432 (1990).
\item Giommi, P. et al., Ap. J., {\bf 378}, 77, (1991).
\item Hara, T. et al., Nucl. Inst. and Meth., {\bf 300} , A332 (1993).
\item Hillas, M., Proc. $19^{th}$ ICRC (La Jolla), {\bf 3}, 445 (1985).
\item Kerrick, A. D. et al., Ap. J, {\bf 438}, L59 (1995).
\item Li, T.-P and Ma, Y.-Q, Ap. J. , {\bf 272}, 317 (1983).
\item Protheroe, R. J., Astron. Express, {\bf 1}, 33 (1984).

\end{list}

\end{document}